\documentstyle[12pt]{article}
\begin{document}
\newcommand{\be}{\begin{equation}}
\newcommand{\ee}{\end{equation}}
\newcommand{\bea}{\begin{eqnarray}}
\newcommand{\eea}{\end{eqnarray}}
\newcommand{\lbl}[1]{\label{eq:#1}}
\newcommand{\ceq}[1]{(\ref{eq:#1})}
\newcommand{\cfig}[1]{\ref{#1}}
\newcommand{\matr}{\left( \begin{array}}
\newcommand{\ematr}{\end{array} \right)}
\newcommand{\dis}{\displaystyle}
\newcommand{\ba}{\begin{array}}
\newcommand{\ea}{\end{array}}
\newcommand{\beqa}{\begin{eqnarray}}
\newcommand{\eeqa}{\end{eqnarray}}
\newcommand{\lsim}{{\;\raise0.3ex\hbox{$<$\kern-0.75em\raise-1.1ex
\hbox{$\sim$}}\;}}
\newcommand{\gsim}{{\;\raise0.3ex\hbox{$>$\kern-0.75em\raise-1.1ex
\hbox{$\sim$}}\;}}
\thispagestyle{empty}
\newcommand{\prstr}{ $e^-e^- \rightarrow
\mu\:\nu\: q\:\bar q \; $ }
\newcommand{\prud}{ $e^-e^- \rightarrow
\mu \:\nu \: d \:\bar u \; $ }
\newcommand{\Del}{\Delta_L^-}
\newcommand{\DDel}{\Delta_L^{--}}

\begin{flushright}
   HIP-1999-56/TH \\
   \end{flushright}
\vspace*{3cm}
\begin{center}
{\bf \large Testing lepton number violation with the reaction $e^-
e^- \rightarrow \mu \:\nu \: q \:\bar q $ }
\end{center}
\vspace*{3cm}
\centerline{J. Maalampi$^a$ and N. Romanenko$^{a,b}$}
\begin{center}
{\footnotesize \it
 $^a$ Theoretical Physics
  Division, Department of Physics,
   University of Helsinki, Finland \\
 $^b$ Petersburg Nuclear Physics Institute,
   Gatchina, Russia}
\end{center}
\vfill \abstract{We investigate the reaction \prstr as a possible
place to test the lepton number violating interactions of singly
charged scalars $\Delta_L^{\pm}$ belonging to a SU(2)$_L$ triplet.
The existence of such scalars is predicted by some majoron models,
as well as by the left-right symmetric extension of the Standard
Model. We show that this reaction may be  observed in $e^-e^-$
collision well below the threshold of  $\Delta_L^{\pm}$. For the
colliding energy $\sqrt{s}=500$ GeV the mass of the singly charged
Higgs triplet may be excluded up to 2 TeV and even more, depending
on values of the appropriate Yukawa couplings. }

 \vspace*{15mm}

{\footnotesize Email addresses: jukka.maalampi@helsinki.fi \\
\hspace*{3cm} nikolai.romanenko@helsinki.fi}
 \vfill
\newpage

Recent experimental results of the SuperKamiokande collaboration
on atmospheric neutrinos \cite{Kam} strongly indicate that
neutrinos oscillate and consequently that at least some of
neutrino species have nonzero mass. This  gives rise to a series
of interesting questions concerning the origin and nature
(Dirac/Majorana) of those masses. Specifically, it would be
interesting and important to know the structure of Higgs sector
responsible for the indicated mass and mixing pattern. For that,
it is obviously  necessary to go beyond the Standard Model (SM) as
in the SM neutrinos are massless. If one does not introduce
right-handed neutrinos, one possibility is to extend the scalar
sector of the SM with an SU(2)$_{\rm L}$ triplet ($\Delta_L$)
Higgs with hypercharge $Y=2$. This could yield  Majorana mass to
left-handed neutrinos through spontaneous symmetry breaking,
assuming that $\langle \Delta_L\rangle \ne 0$.  This is what
happens in some majoron models \cite{Majoron}, and the triplet
Higgses quite naturally appear also in the left-right symmetric
model (LR-model) of electroweak interactions \cite{LR,SS}.

The Higgs triplet $\Delta_L$ with hypercharge $Y=2$ consists of a doubly charged scalar,
a singly charged scalar and
two neutral scalars. The gauge invariance allows the triplet $\Delta_L=(\Delta_L^{0},\Delta_L^{-},
\Delta_L^{--})$ to interact with leptons via a Yukawa coupling that violates the
lepton number $L$ by two units. Due to this property  the search of these particles is  particularly handy and advantageous in electron-electron collisions where one has $L=2$ in the initial state.

The phenomenology of the doubly charged Higgs particle has been
studied, e.g., in \cite{hRll}. It was shown that $\Delta_L^{--}$
with a mass up to 800 GeV may be excluded at the coming LHC
experiments. In the present paper we shall  concentrate on the
search of the singly charged member  of the  Higgs tripet,
$\Delta_L^{-}$ in $e^-e^-$ collisions at a linear collider
\cite{PhysRep}.
 Obviously, due to their different charges, the
phenomenological signatures of the singly charged triplet Higgs
differ considerably from those of the doubly charged Higgs.  The
present experimental data allows to restrict the mass of the
singly charged bosons to be above roughly 100 GeV \cite{PDG}. It
may thus  happen that the singly charged triplet Higgs turns out
to be considerably lighter than its doubly charged counterpart.
Hence it is worthwhile to examine its experimental signatures and
prospects for obtaining information on its properties
independently of the properties of the doubly charged scalar.

There is an important restriction that affects the production and
decay rates of the triplet Higgs $\Delta_L$. The vacuum
expectation value of its neutral member,
$<\Delta_L^0>={v_L}/{\sqrt2}$, is limited to quite small values in
order to avoid a violation of the experimentally well established
relation $\rho \equiv M_W^2/(M_Z^2 \cdot \cos^2 \theta_W)=1$. This
restriction remains the same for the LR-model if the $W_L-W_R$ and
$Z_L-Z_R$ mixings are neglected \cite{Gun1}.
 The present experimental data indicate that $v_L \le 15$ GeV.
Let us note, that  there is an option to extend  the Higgs sector
further so that the equality $\rho=1$ holds at the tree level due
to the so-called "custodial" $SU(2)$ symmetry \cite{MCG}. In this
scenario, which will not be considered in the following, there
would be more than one  charged triplet scalars.

 In this paper we will show that the process
\prstr
provides a  good environment to study  the interactions of the singly charged scalar $\Delta_L^{-}$, quite
independently of the properties of the doubly charged Higgs $\Delta_L^{--}$.
For the anticipated luminosity of electron-electron collider
this process may be observed well below the production threshold of the singly charged Higgs.

 We consider the standard SU(2)$_{\rm L}\times$U(1)$_{\rm Y}$ model with an additional Higgs
 triplet field of hypercharge $Y=2$:
 \be
 {\dis\Delta_L=\matr{cc}\Delta_L^+/ \sqrt{2}&\Delta_L^{++}\\
\Delta_L^0&-\Delta_L^+ /\sqrt2\ematr = (3,1,2)} \ee with the
vacuum expectation value:
\be
\begin{array}{c}  {\dis\langle\Delta_{L}\rangle
=\frac1{\sqrt{2}}\matr{cc}0&0\\v_{L}&0 \ematr}
\end{array}.
\ee This kind of model was first  suggested in \cite{Majoron} in
order to generate Majorana masses for neutrinos, and it also quite
naturally arises as an effective low-energy manifestation of  the
left-right symmetric model \cite{Desh}. The couplings of
$\Delta_L$  with the gauge fields are provided by the usual
kinetic term in the Lagrangian,
\be
L_{\rm kin}=\frac12 {\rm Tr} \left( D_{\mu} \Delta_L \right)^+
\left( D_{\mu} \Delta_L \right), \lbl{kin}
\ee
where
\be
D_{\mu} \Delta_L= \partial_{\mu} \Delta_L + ig'B_{\mu}
\cdot \Delta_L
+\frac{ig}{2} W_{\mu}^a \left[\tau^a,  \Delta_L \right],
\ee
 and its interactions  with fermions are given by a Yukawa coupling of the form
\be
L_{\rm Yuk}=-h_{L,\alpha\beta}\Psi_{\alpha
L}^TC\sigma_2\Delta_L\Psi_{\beta L} \:\:\:+\:{\rm h.c.}, \lbl{Yuk}
\ee where $\Psi$ denotes the lepton doublet $(\nu_l,l^-)$ and
$\alpha$ and $\beta$ are flavour indices. The Yukawa interaction
gives rise to Majorana mass terms
 $m_{\nu_{\alpha\beta}}= h_{L,\alpha\beta}\cdot v_L \;$ for the left-handed
neutrinos.

The interactions of the $\Delta_L$ field described above in
\ceq{kin}, \ceq{Yuk} are the same for both the SM with
additional Higgs triplet and the LR-model.
Of course, the full Lagrangian should also include self-interactions of the
scalar fields, which are governed by the respective scalar potentials.
We will overlook these in the following as a first approximation by assuming no mixing
between the doublet and triplet Higgses and the absence of any further Higgs fields.

The phenomenologically most important feature of the models that
include triplet Higgses is the lepton number non-conservation
arising from the Yukawa coupling \ceq{Yuk}. It makes the
interactions of the triplet Higgses quite complementary to the
Yukawa interactions of the SM Higgs doublet, and it provides  good
opportunities for unambiguous tests of  $\Delta_L^{-}$ production
and decay  in $e^-e^-$ collision experiments. The production of
like-sign charged Higgs bosons via charged vector boson fusion in
several electroweak models was considered in \cite{Rizzo}. It was
shown that cross section strongly depends on the choice of the
Higgs representation and on the parameters of the model. The
models described above were, however, not considered in that
study. It is impossible to distinguish the lepton number
conserving properties of charged Higgs through this kind of
process.

 In \cite{Rizzo} the production of the singly charged Higgses in $e^-e^-$ collisions
was assumed to happen in pairs through a $W^-W^-$ fusion. This
process conserves the lepton number. We are interested, in
contrast, in   processes
 that probe the lepton number violating Yukawa couplings
 \ceq{Yuk}. The pair production, which proceeds through t-channel exchange of
Majorana neutrinos and s-channel exchange of $\Delta_L^{--}$,
 is in this case
not a suitable process to study, however.
 This is because
  the neutrino exchange is proportional to Majorana mass
 of the neutrino and hence is suppressed and the
 $\Delta_L^{--}\Delta_L^-\Delta_L^- $ vertex depends on the
 self-couplings of scalar potential whose values are unknown. We consider instead
a production of a single $\Delta_L^-$ in the process $e^-e^-\to\Delta_L^- W^-_{\mu}$
which proceeds through t-channel exchange
  of Majorana neutrino and s-channel
 exchange of $\Delta_L^{--}$. In this case the
  neutrino exchange is not suppressed as the t-channel neutrino
  has the same chirality in the both vertices. Moreover, in the s-channel process
the strength of the
 $\Delta_L^{--}\Delta_L^-W_{\mu}^-$ vertex does
  not depend on any unknown parameter of the scalar potential
but is determined by the  gauge coupling. The experimentally
clearest final state to study is the one where
  $ \Delta_L^-$ decays to a muon and a muonic neutrino and  $W^-_{\mu}$
decays into two quark jets without missing energy.
  We hence choose the process \prstr for further investigation,
 specifically with the light quarks $d$ and $ \bar{u}$  in final state.

In addition to the two  amplitudes mentioned above, there exist
still another amplitude  contributing in the process \prstr,
namely the one where $\DDel$ is exchanged in s-channel producing
two muons, one of which decaying further into $\nu d\bar{u}$. This
amplitude does not directly depend on the properties of $\Del$ and
is for that reason  undesirable, but its influence is, of course,
unavoidable.

 In our calculations, whose results will be presented in the following, we have
used CompHEP package created in Moscow
 University \cite{CompHEP}. As an input we have used the vertex functions
$\; \sqrt2 \cdot h_{L,ee}\cdot (1-\gamma_5)/2 \;$ and $ \;-g \cdot
(p^{++}_{\mu}-p^-_{\mu})$ for the $e \nu_{e} \Delta_L^+$ and
$W_{L}^- \Delta_L^{++} \Del\;$ couplings, respectively, as well as
 the ordinary electroweak vertices for the couplings of $W^-$ with leptons and
quarks.
 In our calculations we have imposed the following cuts for the final
 state phase space:
 \begin{itemize}
\item[-] Each final state particle has energy greater than 10 GeV
(including neutrino).
\item[-] The transverse energy of each particle (including missing
transverse energy) should be greater than 5 GeV.
\item[-]The opening angle between two quark jets should be more than
$20^o$.
\item[-]Each final state particle should have the outgoing
direction more then $10^o$ away from the beam axis.
\end{itemize}

In Fig. 1 we present the  dependence of the cross section
of the process \prud on the collision energy for the different values of
masses of singly ($M_{\Del}$=100, 400, 700, 1000 GeV)
and doubly charged ($M_{\DDel}=$ 100, 400, 700, 1000 GeV) triplet Higgses.
 The cross sections are dominated by
 the resonance  at $\sqrt{s}=M_{\DDel}$.
  To estimate the width of the peak we have chosen
$\Gamma_{\DDel}=10^{-3} \cdot M$  for the two lepton decays and
the $\DDel \rightarrow \Del W_L^-$ mode was also taken into
account \cite{hRll}.   One may conclude that at 0.01 fb level the
process \prud may be observed away from the $\DDel$ resonance and
even below the $\Del$ threshold.

In Fig. 2 presents the sensitivity of the reaction \prud on the masses of the
 triplet Higgs particles $\Del$ and $\DDel$
 for the collision energy $ \sqrt{s}=500$
 GeV.  We have estimated the values of  the running coupling constants at 500 GeV
 by applying the approximate RG equations of the SM  \cite{Rom}. The influence
of the triplet Higgses on the running, which can be expected to be
quite small, is not taken into account. \mbox{Fig. 2a} displays
the cross section of the \prud process for the different values of
the $\Del$ and $\DDel$ masses, with assuming for the Yukawa
couplings their maximal allowed values that are in accordance with
the present phenomenological constraints \cite{hRll,Maal}: $$
h^2_{ee} < 10^{-5}\cdot M_{\DDel} \; {\rm GeV}, $$
\be
h^2_{\mu \mu} < 10^{-5} \cdot M_{\DDel} \; {\rm GeV.}
 \ee
 If the
mass of doubly charged Higgs is considered to be greater than 100
GeV , then $h_{e e} \cdot h_{\mu \mu} < $ 0.18 or $\sqrt{h_{e e}
\cdot h_{\mu \mu}} <$ 0.4.

As  can be seen from Fig. 2a, the process \prud will be  well
observable at colliding energy 500 GeV for any values of $\Del$
and $\DDel$ masses below 2 TeV if the Yukawa  couplings of the
triplet fields have their maximal allowed values. Furthermore, the
anticipated luminosity of linear collider ($0.03$ fb at $\sqrt s=$
500 GeV \cite{PhysRep}) allows to state that the value of the
product $h_{e e} \cdot h_{\mu \mu}$, which enters as a common
factor in all considered Feynman amplitudes, may be restricted at
least one order of magnitude better than the present bound.
Nevertheless, it is not so easy to separate the influence of the
singly charged Higgs from that of the doubly charged Higgs
particle. First of all let us notice, that in the limit of
$M_{\Del}$ going to infinity, the cross section remains finite due
to contribution of the s-channel  process $e^+e^-\to\DDel\to
\mu^+\mu^-$ followed by subsequent decay of one of the muons to a
$\nu d \bar{u}$. This process does not involve the singly charged
Higgs. If ${\DDel}$  in turn  is decoupled, the amplitude with the
t-channel exchange of Majorana neutrinos would keep the cross
section finite. In order to search for the effects of the singly
charged triplet Higgs we have studied the difference between the
cross sections of the \prud process at a  fixed and an infinite
value (in practice $M_{\Del} = 2$ TeV) of the  $\Del$ mass. In
Fig. 2b we show the dependence of the cross section on the $\DDel$
mass in the case that  $\Del$ is effectively decoupled.
%
Supposing that the mass of $\DDel$ is known one can conservatively
estimate, by setting for the Yukawa coupling $h_{\mu\mu}$ the
largest phenomenologically allowed value, the contribution of the
$\DDel$ mediated processes on the total cross section. When this
is subtracted from the total cross section, what is left is the
contribution of the t-channel neutrino exchange process alone.
This has  a threshold behaviour and its strength gives direct
information on the product ${h_{e e} \cdot h_{\mu \mu}}$ of Yukawa
couplings.

In Fig. 2c we display the 0.03 fb (30 events per year) discovery
contours on the  $\left( M_{\Del},M_{\DDel} \right)-$plane,
 corresponding to the  cross section of the
  isolated t-channel process, for the
different values (0.1, 0.4 and 1.0) of "average" Yukawa couplings
$(h_{\rm Yuk}=\sqrt{h_{e e} \cdot h_{\mu \mu}})$. In the plot the
collision energy is taken as $\sqrt{s}= 500$ GeV.
 It is seen from the figure that the process \prud might
  probe the the mass $M_{\Del}$
  to much larger values than what is
   the production threshold, providing
   that the average Yukawa coupling is
   larger than 0.1 and the collision
     does not happen in the vicinity
     of the $\DDel$ pole. If these
      conditions are not met $\Del$ would
       have detectable effects only when
        it is produced as a real particle.

The constraints ensuing for the average
 Yukawa coupling are presented in Fig. 3
  for different values of the mass
  of the doubly charged Higgs
   ($M_{\DDel}= 400, 1000, 1500 $ and 2000 GeV)
    at $\sqrt{s}=500$ GeV. For a light $\DDel$
     the constraint is tightest,
      about $h_{\rm Yuk}\lsim 0.1$
       and it does not  depend on
         the mass of $\Del$ outside the
        threshold region. The larger $M_{\DDel}$ the weaker
          and also the more dependent on the mass of $\Del$
           is the bound.

  The main SM background to the reaction \prud is due to the process $e^-e^- \rightarrow W^-W^- \nu
  \nu$ studied in \cite{Cuy}. Its cross section
   was estimated to be below  20 fb for
  colliding
   energy $\sqrt{s}\lsim$ 1 TeV. Taking into
   account the branching rations of
      $W$ to the appropriate decay
       modes we can estimate  the background to be about
       \mbox{1 fb}.
        This is typically below the process
   \prud rate, but may be much larger than the singly charged
   contribution to the process. However, reconstructing the
   invariant squared mass of the muon and neutrino pair it would
   be possible to separate background in the cases when the
   mass difference between $\Del$ and $W^-$
is greater than  invariant mass resolution (for $M_{\Del} > 100$
GeV this should be possible). But even in the cases when $M_{\Del}
\simeq M_W$ it is possible to compare the cross sections of \prud
and $e^-e^- \rightarrow d \: \bar{u} \: s \: \bar{c}$ which should
be equal in the SM. Any substantial  difference between these
cross sections would  be a signal of the new physics. In other
words, in order to get rid of the SM background one should
consider the ratio of the cross sections of $e^-e^- \rightarrow d
\; \bar{u} \; s \; \bar{c}$ and \prud.

 In summary, we have shown that the process \prud provides a good test for
 lepton flavor non-conservation of the singly charged scalars. At the
 collision energy 500 GeV the process may be seen well below
 $\Del$ and/or $\DDel$ thresholds for a wide range of the lepton number violating Yukawa
 couplings. The influence of $\Del$ contribution
 (below its threshold) may be
 extracted from the process, if colliding energy
 is away from the  $\DDel$  resonance. The present bounds on the
 Yukawa couplings may be significantly improved.

    \section*{Acknowledgments}

  One of us   (N.R.)  is  grateful to CIMO
  organization for the financial support
  making his stay in Helsinki possible,
 and to the Theoretical Physics Division
  at the Department of Physics of Helsinki University
  for warm hospitality.
  It is also a great pleasure to thank Alexander Pukhov
  for helpful instructions for using the CompHEP package.
This work has been  supported also by the Academy of Finland under
the contract
 40677 and by RFFI grant 98-02-18137.


\newcommand{\bi}{\bibitem}

\end{document}